\documentclass[12pt,a4paper,12pt,superscriptaddress,nofootinbib]{revtex4-1}
\usepackage{graphicx}
\usepackage{amssymb}
\usepackage{afterpage}
\usepackage{colordvi}
\usepackage{epsfig}

\begin{document}

\title{\boldmath The event generator for the 
two-photon process\\ $e^+e^- \to e^+e^-R$ $(J^{PC}=0^{-+})$
in the single-tag mode}

\author{V.~P.~Druzhinin}
\affiliation{Budker Institute of Nuclear Physics, Novosibirsk 630090, Russia }
\affiliation{Novosibirsk State University, Novosibirsk 630090, Russia }
\author{L.~A.~Kardapoltsev}
\affiliation{Novosibirsk State University, Novosibirsk 630090, Russia }
\affiliation{Budker Institute of Nuclear Physics, Novosibirsk 630090, Russia }
\author{V.~A.~Tayursky}
\email{tayursky@inp.nsk.su}
\affiliation{Budker Institute of Nuclear Physics, Novosibirsk 630090, Russia }

\begin{abstract}
{The Monte Carlo event generator GGRESRC is described.
The generator is developed for
simulation of events of the two-photon
process $e^+e^-\to e^+e^- R$, where R is a pseudoscalar
resonance, $\pi^0 $, $ \eta $, $ \eta^\prime $, $ \eta_c $, or $ \eta_b $.
The program is optimized  for generation of two-photon
events in the single-tag mode. For single-tag events, radiative correction
simulation is implemented in the generator including photon emission
from the initial and final states.}
\end{abstract}

\maketitle
\setcounter{footnote}{0}

\section {Introduction}
The purpose of this work is to develop an efficient event generator for 
the process of the two-photon resonance production 
$e^+e^-\to e^+e^- R$ in the so-called single-tag mode, when one of
the final electrons\footnote{Unless otherwise
specified, we use the term ''electron'' for either an electron or a
positron.} is scattered at a large angle and detected. 
Such generator is needed for simulation of experiments on the measurement 
of the meson-photon transition form factors. 
The generator GGRESRC described in this work was used
for the measurement of the transition form factors for the $\pi^0$, $\eta$,
$\eta^\prime$, and $\eta_c$ mesons with the BABAR detector.
To achieve required accuracy ($\sim 1\%$), the radiative corrections 
to the Born cross section are taken into account. In particular, 
extra photon emission from the initial and final states are simulated. 

\begin{figure}[t!]
\includegraphics[width=.45\textwidth]{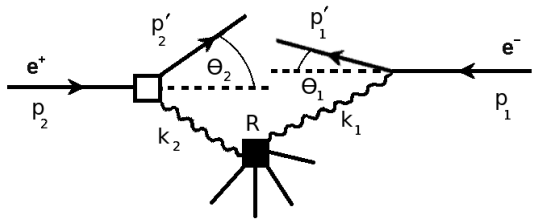}
\caption{\label{fig1}
The diagram of the two-photon process $e^+ e^- \to e^+ e^- + R $.}
\end{figure}

In the two-photon process $e^+ e^- \to e^+ e^-R $,
the virtual photons, radiated by the colliding electrons,
form
a $C$-even resonance with the four-momentum 
${k} = {k} _1 +{k} _2 $ (see Fig.~\ref{fig1}).
Let $Q_2^2 $ be the absolute value of the four-momentum squared,
carried by the space-like photon connected with the tagged (detected) electron , 
while $Q_1^2$ be the same parameter for the untagged (undetected) electron
($Q_1^2\approx 0$). The transition form factor is determined from
the measured differential cross section 
$({\rm d}\sigma/{\rm d}Q_2^2)_{\rm data}$ and
the MC calculated cross section 
$({\rm d}\sigma/{\rm d}Q_2^2)_{\rm MC}$:
\begin{equation}
|{F}^{\rm data}_{\gamma^* \gamma R}(Q_2^2)|^2=
\frac{({\rm d}\sigma/{\rm d}Q_2^2)_{\rm data}}
{({\rm d}\sigma/{\rm d}Q_2^2)_{\rm MC}}
|{F}^{\rm MC}_{\gamma^* \gamma R}(Q_2^2)|^2,
\end{equation}
where $|{F}^{MC}_ {\gamma^*\gamma R} (Q_2^2)|^2$
is the transition form factor used in MC simulation.

\section {Born cross section}
To describe the process $e^+ e^- \to e^+ e^- R $ we use the notations 
defined in Fig.~\ref{fig1}, and the following six invariants:
\begin{eqnarray}
&t_1=-Q_1^2={k}_1^2, \quad t_2=-Q_2^2={k}_2^2,\nonumber\\
&s_1=({p}_1'+{k})^2, \quad s_2=({p}_2'+{k})^ 2,
\label{eq2}\\
&s=({p}_1+{p}_2)^2, \quad W^2={k}^2=({k}_1+{k}_2)^2.\nonumber
\end{eqnarray}
The differential cross-section for this process
in the lowest QED order is given by~\cite{BGMS}:
\begin{equation}
{\rm d}\sigma=\frac{\alpha^2}{16\pi^4t_1t_2}\sqrt{\frac{({k}_1
{k}_2)^2-t_1 t_2}{({p}_1 {p}_2)^2-m_e^4}}\Sigma 
\frac{{\rm d}^3\vec{p'}_1}{E'_1}
\frac{{\rm d}^3\vec{p'}_2}{E'_2},
\label{eq3}
\end{equation}
where $\alpha$ is the fine structure constant, 
$m_e $ is the electron mass, $E'_i$ ($i$=1,2) are the energies of 
the scattered electrons and 
\begin{eqnarray}
\Sigma& = &4 \rho^{++}_1 \rho^{++}_2 \sigma_{TT} +
2 \rho^{++}_1  \rho^{00}_2 \sigma_{TS} +
 2\rho^{00}_1  \rho^{++}_2 \sigma_{ST} +
\rho^{00}_1  \rho^{00}_2 \sigma_{SS}
\label{eq4}\\
&+&2|\rho^{+-}_1 \rho^{+-}_2| \tau_{TT} \cos\,2\tilde{\phi}
-8|\rho^{+0}_1 \rho^{+0}_2| \tau_{TS} \cos\,\tilde{\phi}. 
\nonumber
\end{eqnarray} 
Here $\tilde{\phi}$ is the angle between the electron and positron scattering 
planes
in the center-of-mass (c.m.) frame of the virtual photons, 
$\sigma_{ab}$ are the $\gamma^\ast\gamma^\ast \to R$ cross sections 
for unpolarized transverse ($a, b=T$) and scalar ($a, b=S$) photons.
The interference terms containing the functions $\tau_{ab}$ arise due to 
virtual photon
polarization. The function $\tau_{TT}$ is the difference between cross 
sections for 
transverse photons with the parallel and orthogonal linear polarizations: 
$\tau_{TT}=\sigma_\parallel-\sigma_\perp$, while the cross section for 
unpolarized
photons is $\sigma_{TT}=(\sigma_\parallel+\sigma_\perp)/2$.

The effects of the strong interaction are completely contained in the 
functions $\sigma_{ab}$ and $\tau_{ab}$. All other functions entering in
Eq.~(\ref{eq4}) are calculable with QED. The expressions for the 
virtual photon density matrices  
$\rho^{++}_i$, $\rho^{+-}_i$, $\rho^{+0}_i$, $\rho^{00}_i$ ($i=1,2$) 
can be found in Ref.~\cite{BGMS}.

In the case of the pseudoscalar meson production,
only the functions $\sigma_{TT}$ and $\tau_{TT}$ are non-zero, and 
$\tau_{TT}=-2\sigma_{TT}$~\cite{poppe}. The cross section $\sigma_{TT}$ 
for a narrow
pseudoscalar meson with the mass $M_R$ can be
written in term of the transition form factor:
\begin{equation}
\sigma_{TT}(W,Q_1^2,Q_2^2)=8\pi\frac{\Gamma_{\gamma\gamma}}{M_R}
\left |\frac{{F}(Q_1^2,Q_2^2)}{{F}(0,0)} \right |^2,
\,\,|{F}(0,0)|^2=\frac{4\Gamma_{\gamma\gamma}}{\pi\alpha^2 M_R^3},
\label{eq5}
\end{equation}
where $\Gamma_{\gamma\gamma}$ is the meson two-photon width.
It should be noted that some two-photon event generators neglect the term
with $\tau_{TT}$. This approach may be reasonable only for study of two-photon 
processes
in the no-tag mode, when both the electrons are scattered at small angles.
The $\tau_{TT}$ term gives a sizable contribution
to the differential cross section ${\rm d}\sigma/{\rm d}Q_2^2$ at large $Q_2^2$ 
and should be taken into account in simulation of single-tag experiments. 

In the GGRESRC events generator we perform integration of the differential 
cross section
using invariant variables (\ref{eq2}). For a narrow pseudoscalar resonance,  
Eq.~(\ref{eq3}) can be rewritten:
\begin{equation}
{\rm d}\sigma=\frac{4\alpha^2\Gamma_{\gamma \gamma}}
{\pi s^2 t_1^2 t_2^2 M_R^3}
\left|\frac{F(t_1,t_2)}{F(0,0)}\right |^2 B\frac{{\rm d}t_2  {\rm d}t_1 
{\rm d}s_1 {\rm d}s_2}{\sqrt{-\Delta_4}},
\label{eq6}
\end{equation}
where
 $\Delta_4(s,s_1,s_2,t_1,t_2,M_R^2,m_e^2)$ is the Gram determinant~\cite {BK}.
The physical region in the variables $s_1$, $s_2$, $t_1$, $t_2$ is defined by 
the condition $\Delta_4 \le 0 $.
The function $B$ coincides, up to a factor, with the function $\Sigma$ 
(Eq.~(\ref{eq4})) for pseudoscalar mesons. It was calculated in Ref.~\cite {BKT} 
and is given by
\begin{equation}
B=\frac{1}{4}t_1t_2B_1-4B_2^2+m_e^2B_3,
\label{eq7}
\end{equation}
where
\begin{eqnarray}
B_1&=&(4{p_1}{p_2}-2{p_1}{k_2}-2{p_2}{k_1}+{k_1} {k_2})^2+
({k_1}{k_2})^2-16t_1t_2-16m_e^4,\nonumber\\
B_2&=&({p_1}{p_2})({k_1}{k_2})-({p_1}{k_2})({p_2}{k_1}),\label{eq8}\\
B_3&=&t_1(2{p_1}{k_2}-{k_1}{k_2})^2+
t_2(2{p_2}{k_1}-{k_1}{k_2})^2+4m_e^2({k_1}{k_2})^2,\nonumber
\end{eqnarray}

\begin{figure}[t!]
\includegraphics[width=.6\textwidth]{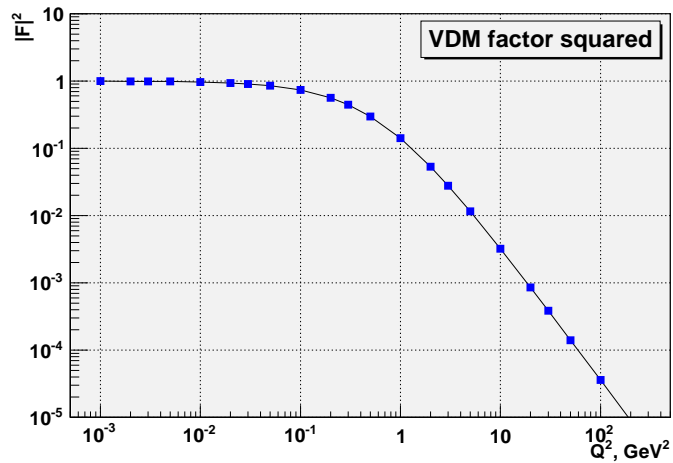}
\caption{\label{fig2}
The $Q^2$ dependence of the form factor $|F|^2$
at $Q_1^2=0$, $\Lambda=m_{\rho}$=0.7755 GeV.}
\end{figure}

To describe the $Q_1^2$ and $Q_2^2$ dependencies of the transition form 
factor 
$F(Q_1^2,Q_2^2)$, the two options are implemented in the generator: 
$F(Q_1^2,Q_2^2)=F(0,0)$, 
and the vector-dominance model (VDM)
\begin{equation}
|F|^2=\frac{1}{(1+Q^2_1/\Lambda^2)^2(1+Q^2_2/\Lambda^2)^2},
\label{eq9}
\end{equation}
where $\Lambda=m_\rho$ for $\pi^0$, $\eta$, $\eta'$ production,
$\Lambda=m_{J/\psi}$ for $\eta_c$, and
$\Lambda=m_\Upsilon$ for $\eta_b$. 
The $Q_2^2$ dependence of the $|F|^2 $  calculated with 
Eq.~(\ref {eq9}) at $Q_1^2=0 $ for $\Lambda=m_{\rho} $,
is shown in Fig.~\ref{fig2}.

Four-dimensional Monte-Carlo integration of Eq.~(\ref{eq6}) is performed using 
the method developed for the GALUGA two-photon event generator~\cite{S}.
In this method, in particular, the 
invariant variables are generated in the order 
$t_2$, $t_1$,  $s_1 $, $s_2 $.
This allows to set a restriction on $Q_2^2$ at the beginning of 
the event generation and significantly increase the generation efficiency for 
single-tag events. The values  of the generated invariants, 
are then used together with a random azimuthal angle of the system 
of the final 
particles to calculate the 4-momenta of the scattered electron, positron, 
and produced resonance. The formulae to do this can be found in Ref.~\cite{T}. 
The
main decay modes for $\pi^0$, $\eta$, and $\eta^\prime$ are also simulated 
according to Ref.~\cite{T}.

The total widths of the $\eta_c$ and $\eta_b$ resonances are comparable 
or even larger than the mass resolution of modern detectors.
Therefore, 
the mass distributions for 
these resonances are generated using Breit-Wigner distributions. 

\section {Radiative correction}
In the no-tag mode, when both the electron and the positron are scattered 
predominantly 
at small angles, the radiative correction to the Born cross section is 
expected to be 
small, less than 1\%~\cite{RC}. The situation changes 
drastically in the single-tag mode, at a large electron scattering angle. 
At large $Q^2$
the correction due to extra photon emission from the initial state may 
reach several
percents and should be taken into account in simulation.  

\begin{figure}[t!]
\includegraphics[width=.85\textwidth]{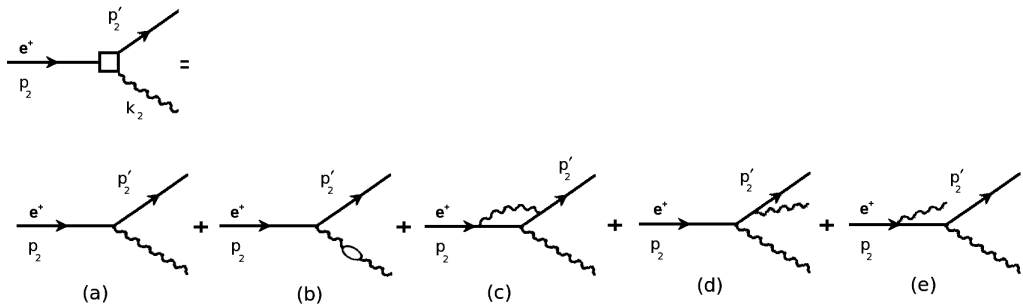}
\caption{\label{fig3}
Diagrams used for calculation of the radiative correction.}
\end{figure}

The process-independent formula for the radiative correction in 
the next-to-leading
order for two-photon processes in the single-tag mode 
was obtained in Ref.~\cite {OK}. The main contribution to the correction 
comes from
the vertex of the tagged electron. The corresponding contribution of the 
untagged-electron vertex is expected to be smaller than 0.5\% and neglected.
Fig.~\ref{fig3} shows the diagrams taken into account in Ref.~\cite {OK}. They
substitutes for the left-hand vertex in Fig.~\ref{fig1}. 

The cross section for a single-tag experiment is given by: 
\begin{equation}
{\rm d}\sigma  =   
{\rm d}\sigma_B(1+\delta)={\rm d}\sigma_B(1+\delta^\prime+\delta_{VP}),
\label{eq10}
\end{equation}
where ${\rm d}\sigma_B$ is the lowest-order cross section for the two-photon
process given, for example, by Eq.~(\ref{eq6}). The total radiative correction
is separated into two parts:
\begin{enumerate}
\renewcommand{\theenumi}{\roman{enumi}}
\item $\delta^\prime$, which includes the virtual correction
due to the interference between the diagrams (a) and (c),
soft-photon part of diagrams (d)+(e), and the corrections
due to real photon emission from the initial (diagram (e)) and final
(diagram (d)) states,
\item $\delta_{VP}$, the vacuum polarization correction due to
the interference between the diagrams (a) and (b).
\end{enumerate}
To obtain $\delta^\prime$ we have used the result of Ref.~\cite{OK} 
for the total radiative correction, removing from it the contribution 
of the vacuum polarization diagram, $\delta_e$ (in Ref.~\cite{OK}
only electron contribution was taken into account). 
The resulting $\delta^\prime$ is given by
\begin {equation}
\delta^\prime=-\frac{\alpha}{\pi}
\Biggl \{
\biggl [
\ln\frac{1}{r_{max}}-\frac{3}{4}
\biggr]
(L-1)+\frac{1}{4}
\Biggr\}.
\label{eq11}
\end{equation}
where $r_{max}$ ($\ll 1$) is the maximum energy of the photon emitted from 
the initial 
state in units of the beam energy $E_b$,
$L =\ln {(Q^2/m_e^2)}$, and $Q^2$ is the absolute value of the momentum
transfer squared to the electron. The formula does not contain any 
restriction on
the energy of the photon emitted from the final state,
i.e. the cross section  given by Eq.~(\ref{eq10}) is calculated for the case 
when the 
tagged electron is allowed to radiate a photon of any possible energy.
The values of the correction $\delta^\prime$ for nine representative
sets of $Q^2$ and ${r_{max}}$ are listed in Table~\ref{tab1}. 
\begin{table}[htb]
\begin{center}
\caption{\label{tab1} The correction $\delta^\prime$ (\%) 
for the various values of 
$r_{max}$ and $Q^2$.}
\begin{tabular}{|c|c|c|c|}
\hline
$Q^2$ (GeV$^2$) & $r_{max}$=0.03 & $r_{max}$=0.05 & $r_{max}$=0.1 \\
\hline
 1   & -9.1 & -7.4 & -5.2 \\
 10  & -10.6 & -8.6 & -6.0 \\
 100 & -12.1 & -9.8 & -6.8 \\
\hline
\end {tabular}
\end {center} 
\end {table}  
In the $Q^2$ region from 1 to 100 GeV$^2$ available for experiments
at $B$-factories, the correction reach 5--7\% even
with the relatively loose restriction  ($r_{max}=0.1$) on the scaled energy
of the undetected photon emitted from the initial state. 

\begin{figure}[t!]
\includegraphics[width=.65\textwidth]{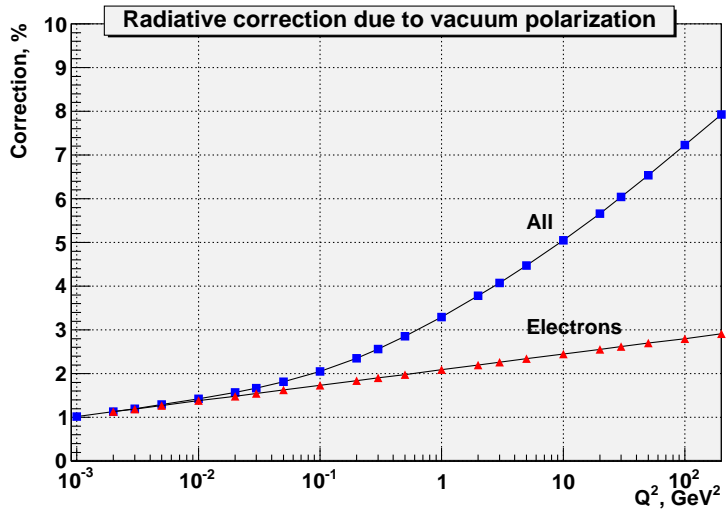}
\caption{\label{fig4}
The vacuum polarization correction as a function of $Q^2$. 
The curve ''All'' shows $\delta_{VP}$ calculated in Ref.~\cite{CMD-2} with
account of contributions from $e$, $\mu$, $\tau$, and hadrons.
The curve ''Electrons'' -- represents the contribution only 
from electrons, $\delta_e$.}
\end{figure}

The correction $\delta^\prime$ is partly compensated by the vacuum 
polarization correction $\delta_{VP}$, for which 
we use the results of Ref.~\cite{CMD-2}, which includes the 
contributions from
the $e$, $\mu$, $\tau$ leptons, and hadrons. The $Q^2$ dependence of $\delta_{VP}$ 
is shown in Fig.~\ref{fig4} in comparison with $\delta_e$.
The values of the total correction $\delta = \delta ' + \delta _ {VP}$
calculated for for nine representative
sets of $Q^2$ and ${r_{max}}$ are listed in Table~\ref{tab2}.
\begin{table}[htb]
\begin{center}
\caption{\label{tab2} Total radiative correction $\delta=\delta^\prime+\delta_{VP}$.}
\begin{tabular}{|c|c|c|c|}
\hline
$Q^2$ (GeV $^2$) & $r_{max}$=0.03 & $r_{max}$=0.05 & $r_{max}$=0.1 \\
\hline
 1 & -5.9 & -4.3 & -2.0 \\
 10 & -5.6 & -3.7 & -1.0 \\
 100 & -4.8 & -2.6 & +0.4 \\
\hline
\end {tabular}
\end {center} 
\end {table}  

The emission of the hard photon by the electron distorts the kinematics 
of two-photon event. To model how this effect influences the detection 
efficiency, 
the event generator includes generation of extra photons emitted from the 
initial and final states. 

\subsection {Simulation of initial state radiation}
For simulation of the initial state radiation (ISR), 
it is convenient to represent the radiative correction in the form 
\begin{equation}
1+\delta'\approx
\left[1+\frac{\alpha}{\pi} \left(\frac{3}{4}L-1\right)\right] 
\int_{0}^{r_{max}} \frac{\beta dr}{r^{1-\beta}},
\label{eq12}
\end {equation}
where $\beta=(\alpha/\pi)(L-1)$, $r=E_\gamma/E_b$, and $E_\gamma$ 
is the energy of the ISR photon.

The function under the integral can be interpreted as the energy spectrum 
for photons radiated from the initial state.
Indeed, at $Q^2=1\div 100$ GeV$^2$ the parameter $\beta$ is small 
($\beta=0.033\div 0.044$), and this function coincides approximately 
with the energy spectrum for hard photons, radiated from the 
initial state~\cite {OK}:
\begin {equation}
\frac{{\rm d}N}{{\rm d}r} = \frac{\alpha (L-1)}{\pi r}. 
\end {equation}     

For simulation of the extra photon emission, we replace the 
four-dimensional integration in 
Eq.~(\ref{eq6}) to five-dimensional one with $r$ as the outermost integration
variable
\begin{equation}
{\rm d}\sigma=\left[1+\frac{\alpha}{\pi} \left(\frac{3}{4}L-1\right)\right]
\frac{\beta}{r^{1-\beta}}{\rm d}\sigma_B{\rm d}r
\label{eq14}
\end{equation}
The vacuum polarization correction is included by the substitution  
\begin{equation}
\alpha^2 \to \alpha^2 (1 +\delta_{VP}(Q_1^2))(1 +\delta_{VP}(Q_2^2))
\label{eq15}
\end{equation}
in the Born cross section ${\rm d}\sigma_B$.

In simulation of the initial state radiation, the approximation is used that
the photon is emitted strictly along the initial direction of the 
radiating
electron. Since the energy of the photon is restricted by the condition
$r<r_{max}$, we expect that this approximation does not lead to a significant 
systematics in determination of the detection efficiency. Note
that selection criteria used in data analysis should provide the fulfillment 
of the condition $r<r_{max}$ for both experimental and simulated events.

To increase simulation efficiency, the variable $r$ is initially generated 
according to the $\beta_0/r^{1-\beta_0}$ distribution with 
$\beta_0 =\beta(Q^2_{min})$, where $Q^2_{min}$ is a lower bound on
the tagged-electron $Q^2$ for simulated single-tag event.
If the generated value of $r$ is higher than a threshold $r_{min}$,
the photon is added to the list of final particles in an event.
The scattered $e^+$ and $e^-$, and the pseudoscalar meson are then generated 
in the frame with the shifted c.m. energy of $2E_b\sqrt{1-r}$.
If $r<r_{min}$, the photon is not generated, and the c.m. energy is not 
shifted, but the radiative correction factor in the cross section 
(see Eq.~(\ref{eq14})) is calculated.

\subsection{Simulation of final state radiation}
\begin{figure}[hbt!]
\includegraphics[width=.55\textwidth]{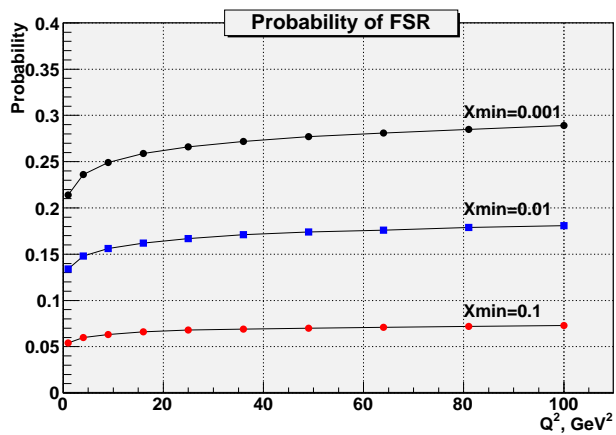}
\caption{\label{fig5}
The $Q^2$ dependence of the final state radiation probability.}
\end{figure}

The final state radiation (FSR) is simulated after the generation of 
the two-photon event. The final electron scattered at a large angle 
is ``decayed'' 
to  $e+\gamma$ with some probability. The final-meson four-momentum is then 
modified to provide the energy and momentum balance.
The probability of the emission of the photon with the energy greater 
than $E_{\gamma,min}$ equals
\begin {equation}
P(Q^2,x_{min})=
\frac {\alpha} {\pi (1 +\delta')}
\biggl [
(L-1) \ln\frac {1} {x _ {min}} -\frac {3} {4} L+1
\biggr],
\label{eq16}
\end {equation}
where $x_{min}=E_{\gamma,min}/E$, and $E$ is the electron energy before
FSR simulation.
This formula is obtained by integration of the FSR photon spectrum given
by Eq.~(23) of Ref.~\cite {OCK}. 
The $Q^2$ dependence of the FSR probability calculated for 
$x_{min}=0.1$, 0.01 and 0.001 is shown in Fig.~\ref{fig5}.

The photon energy $E_\gamma $ and angle $\theta_\gamma$ with
respect to the electron direction before radiation are
generated according to the following distribution function~\cite{OCK}:  
\begin{equation}
\frac {{\rm d}N} {{\rm d}x {\rm d}\cos \theta_\gamma} =
\frac {\alpha} {\pi x} \frac{1-x+x^2/2} {1-\beta \cos \theta_\gamma}, \quad
\label{eq17}
\end{equation}
where $x=E_\gamma/E$, $\beta =\sqrt{1-m_e^2/E^{\prime 2}}$, and $E^\prime$ is
the electron energy after the photon emission.
 
\section {Comparison with other generators}

The comparison of the total cross sections in the no-tag mode 
obtained with GGRESRC and the two other generators of two-photon
events, GGRESPS~\cite{T} and TWOGAM~\cite{TWOGAM}, was performed.
The results of Monte-Carlo calculations are identical for all the 
three generators, if the mass of the meson, its two-photon width, 
and $Q^2$-dependence of the form factor are set to be the same 
in the generators. The GGRESRC and
GGRESPS  use the same formula (Eq.~(\ref{eq6})), but
different orders of integration over the invariant variables.
The TWOGAM generator was developed for the CLEO measurements of the 
meson-photon transition form factors~\cite{CLEO-2}. It is based
on the BGMS formalism~\cite{BGMS} (see Eq.~(\ref{eq3})) and uses
the completely different integration variables, the momenta of the
final electrons.

For GGRESRC in the regime without radiative corrections and TWOGAM,
the comparison of the $Q^2$ spectra, obtained for the process of the 
$\pi^0$ production in the single-tag mode, was performed. The spectra was 
found to be in agreement within the Monte-Carlo statistical errors.
\section {Generator parameters}
The parameters of the event generator are listed in Table~\ref{tab3}.
The recommended values for the parameters Rmax, Rmin, and Kmin
are given in brackets.
To simulate no-tag events, the parameters Q1Smin, Q2Smin, and IRad
should be set to zero.
The regime with radiative correction (IRad=1) is used only in the single-tag
mode.
\begin{table}[htb]
\begin{center}
\caption{\label{tab3} Parameters of the generator GGRESRC.}
\begin{tabular}{|l|l |}
\hline
Name & Description \\
\hline
{ \bf Eb} & beam energy (GeV) \\
{ \bf IR} & produced meson: $\pi^0$ ($=1$), $\eta$ ($=2$), $\eta^\prime$ ($=3$),
$\eta_c$ ($=4$), $\eta_b$ ($=5$) \\
{ \bf IMode} & meson decay mode (see Table~\ref{tab4}) \\ 
{ \bf KVMDM} & form factor model: constant ($=0$), VDM ($=1$) \\
{ \bf Itag} & tagged particle: positron ($=1$), electron ($=2$), mix ($=3$)\\ 
{ \bf IRad} & simulation with/without radiative correction ($=1/0$) \\
{ \bf Rmax} & maximal energy of ISR photon in units of Eb ($ 0.1 $) \\
{ \bf Rmin} & minimal energy of ISR photon in units of Eb ($ 10 ^ {-4} $) \\
{ \bf Kmin} & minimal energy of the FSR photon ( 0.001 GeV ) \\
{ \bf Q1Smin} & minimal momentum transfer squared to the untagged electron \\
{ \bf Q1Smax} & maximal momentum transfer squared to the untagged electron \\
{ \bf Q2Smin} & minimal momentum transfer squared to the tagged electron \\
{ \bf Q2Smax} & maximal momentum transfer squared to the tagged electron  \\
{ \bf Fmax} & maximum weight of events \\
\hline
\end {tabular}
\end {center} 
\end {table}

The resonances decay modes implemented in the generator
are listed in Table 4. The decay models used are
described in Ref.~\cite{T}. 
If parameter IMode equals 0, the meson decay is not simulated.
 
\begin{table}[htb]
\begin{center}
\caption{\label{tab4} The meson decay modes in GGRESRC. 
If IMode=0, the meson decay is not simulated.}
\begin{tabular}{|c|c|l|c |}
\hline
Meson     & IMode & Decay channel & Branching                   \\  
          &       &               &  fraction~\cite {PDG}  (\%) \\
\hline
 $\pi^0 $ & 1& $ \pi^0 \to 2\gamma $ & 98.798 \\  
          & 2& $ \pi^0 \to e ^ + e ^-\gamma $ & 1.198 \\  
\hline
          & 1& $ \eta \to 2\gamma $ & 39.31 \\
$ \eta $ & 2& $ \eta \to 3\pi^0, \quad \pi^0 \to 2\gamma $ & 31.4 \\ 
         &3& $ \eta \to \pi ^ + \pi ^- \pi^0, \quad \pi^0 \to 2\gamma $ &22.457 \\  
         &4& $ \eta \to \pi ^ + \pi ^- \gamma $ & 4.6 \\  
\hline
         &1& $ \eta ' \to 2\gamma $ & 2.1 \\
$ \eta ' $ & 2& $ \eta ' \to \pi^ + \pi^- \eta, \quad \eta \to 2\gamma$&17.532 \\
        &3& $ \eta ' \to \pi ^ + \pi ^- \eta, \quad \eta \to 3\pi^0,\quad
\pi^0 \to 2\gamma $ & 14.004 \\
 &4& $ \eta ' \to \pi ^ + \pi ^- \eta, \quad \eta \to \pi ^ + \pi ^-
\pi^0, \quad \pi^0 \to 2\gamma $ & 10.016 \\  
 &5& $ \eta ' \to \pi ^ + \pi ^- \eta, \quad \eta \to \pi ^ + \pi ^-
\gamma $ & 2.0516 \\
 &6& $ \eta ' \to 2\pi^0 \eta, \quad \eta \to 2\gamma, \quad \pi^0 
\to 2\gamma $ & 7.943 \\
 &7& $ \eta ' \to 2\pi^0 \eta, \quad \eta \to 3\pi^0, \quad \pi^0 
\to 2\gamma $ & 6.3445 \\
 &8& $ \eta ' \to 2\pi^0 \eta, \quad \eta \to \pi ^ + \pi ^- \pi^0, 
\quad \pi^0 \to 2\gamma $ & 4.5375 \\
 &9& $ \eta ' \to 2\pi^0 \eta, \quad \eta \to \pi^+\pi^-\gamma, \quad
\pi^0 \to 2\gamma $  & 0.9294 \\  
 &10& $ \eta ' \to \rho^0 \gamma, \quad \rho^0 \to \pi^+ \pi^-$& 29.4 \\
\hline
$ \eta_c $ & 1& $ \eta_c \to K_SK ^ +\pi ^- $ + c.c. & 2.33 \\
           & 2& $ \eta_c \to 2\gamma $ & 0.024 \\
\hline
$ \eta_b $ & 1& $ \eta_b \to 2\gamma $ & - \\
\hline
\end {tabular} 
\end {center}
\end {table}

In general terms the GGRESRC simulation algorithm is the following:
\begin {itemize}
\item The electron and positron collide in the c.m. frame ($S_0 $). 
In this frame the positive
$z$-axis is defined to coincide with the $e^-$ beam direction.
\item The emission of a hard photon from the initial state is simulated. 
The photon is emitted along the collision axis.
If the ISR photon energy is greater than
$r_{min}E_b$, the photon momentum is stored in the list of final particles.
\item The scattered electrons and the resonance are generated in the new c.m.   
frame ($S_1$) with the c.m. energy $2E_b\sqrt{1-E_\gamma/E_b}$;
$S_1=S_0 $ if $E_\gamma/E_b<r_{min}$.
\item The final state radiation is simulated. If the photon energy
is greater than $k_{min}$, its parameters are stored in the list of final 
particles. The momenta of the tagged electron and the produced meson 
are modified.
\item The meson decay is simulated.
\item The momenta of the final particles are transformed from  
$S_1$ to  $S_0$ frame.
\end{itemize}
When required statistics is collected, the total 
cross section for the two-photon process with
radiative corrections is calculated and printed.

\begin {table} [htb]
\begin {center}
\caption{\label{tab7} The simulation parameters 
used for calculation of the distributions shown in 
Figs.~\ref{fig7}--~\ref{fig14}.}
\begin{tabular}{|l|c|l |}
\hline
Parameter & Value & Comment \\
\hline
Eb      & 5.29  & beam energy (GeV) \\
IR      & 1    & meson: $\pi^0$ \\
IMode   & 1    & decay mode: $\pi^0 \to 2\gamma$ \\
KVMDM   & 1    & VDM form factor (Eq.~(\ref{eq9})) is used \\
Itag    & 1    & the final positron is tagged\\
IRad    & 1    & radiative corrections are simulated\\
Rmax    & 0.1  & maximal energy of ISR photons in units of $E_b$ \\
Rmin    & 10$^{-4}$& minimal energy of ISR photons in units of $E_b$ \\
Kmin    & 0.001 & minimal energy of FSR photons (GeV) \\
Q1Smin  & 0    & $Q^2_{min}$ for $e^-$ (GeV$^2$) \\
Q1Smax  & 1.5  & $Q^2_{max}$ for $e^-$ (GeV$^2$) \\
Q2Smin  & 1.5  & $Q^2_{min}$ for $e^+$ (GeV$^2$) \\
Q2Smax  & 9    & $Q^2_{max}$ for $e^+$ (GeV$^2$) \\
\hline
\end {tabular}
\end {center}
\end {table}

\section {Example of $e^+ e^- \to e^+ e^- \pi^0$ simulation}
In this section, some distributions 
for the process $e^+ e^- \to e^+ e^- + \pi^0$, $(\pi^0 \to 2\gamma)$ 
obtained with the generator GGRESRC, are presented.
In Table~\ref{tab7} the parameters of the generator used in the simulation 
are listed.

At these parameter values, 57\% of events do not contain extra photons,
22\%, 28\%, and 6\% of events contain ISR photon, FSR photon, 
and both ISR and FSR photons, respectively. The obtained 
cross section of the process and average radiative correction are:
$\sigma=$0.99 pb, $\delta=$-0.6\%.

\begin{figure}[hbt!]
\includegraphics[width=.55\textwidth]{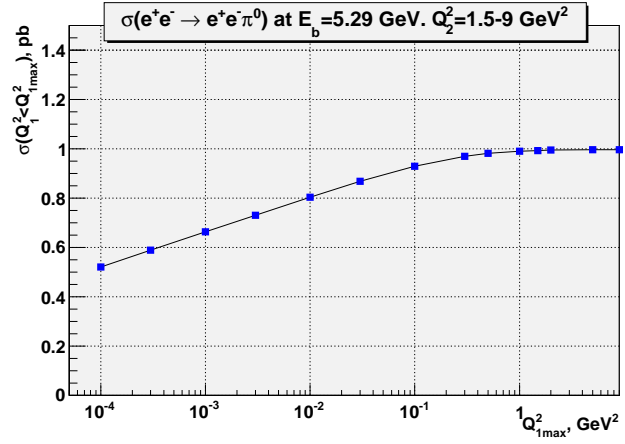}
\caption{\label{fig7}
The $e^+ e^- \to e^+ e^- \pi^0$ cross section
as a function of the limit on $Q_1^2$.}
\end{figure}

The calculated cross section as a function of the restriction on 
$Q_1^2$ (the values of the other parameter are equal to those in 
Table~\ref{tab7}) 
is shown in Fig.~\ref{fig7}.
One can see that at $Q_{1max}^2 \approx$ 1.5 GeV$^2$ the cross section
reaches an asymptotic value.  

\begin{figure}[hbt!]
\includegraphics[width=.45\textwidth]{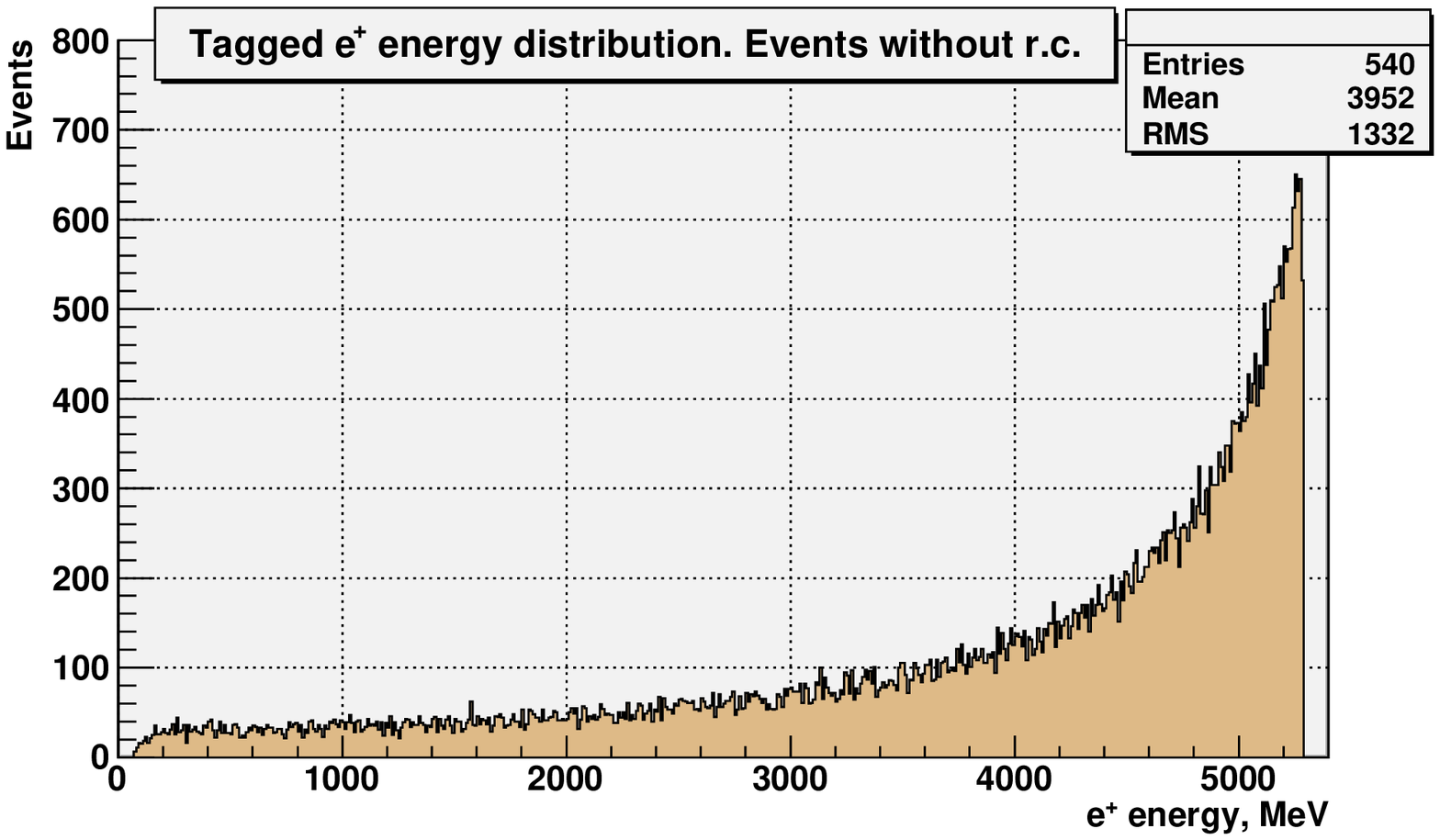}
\includegraphics[width=.45\textwidth]{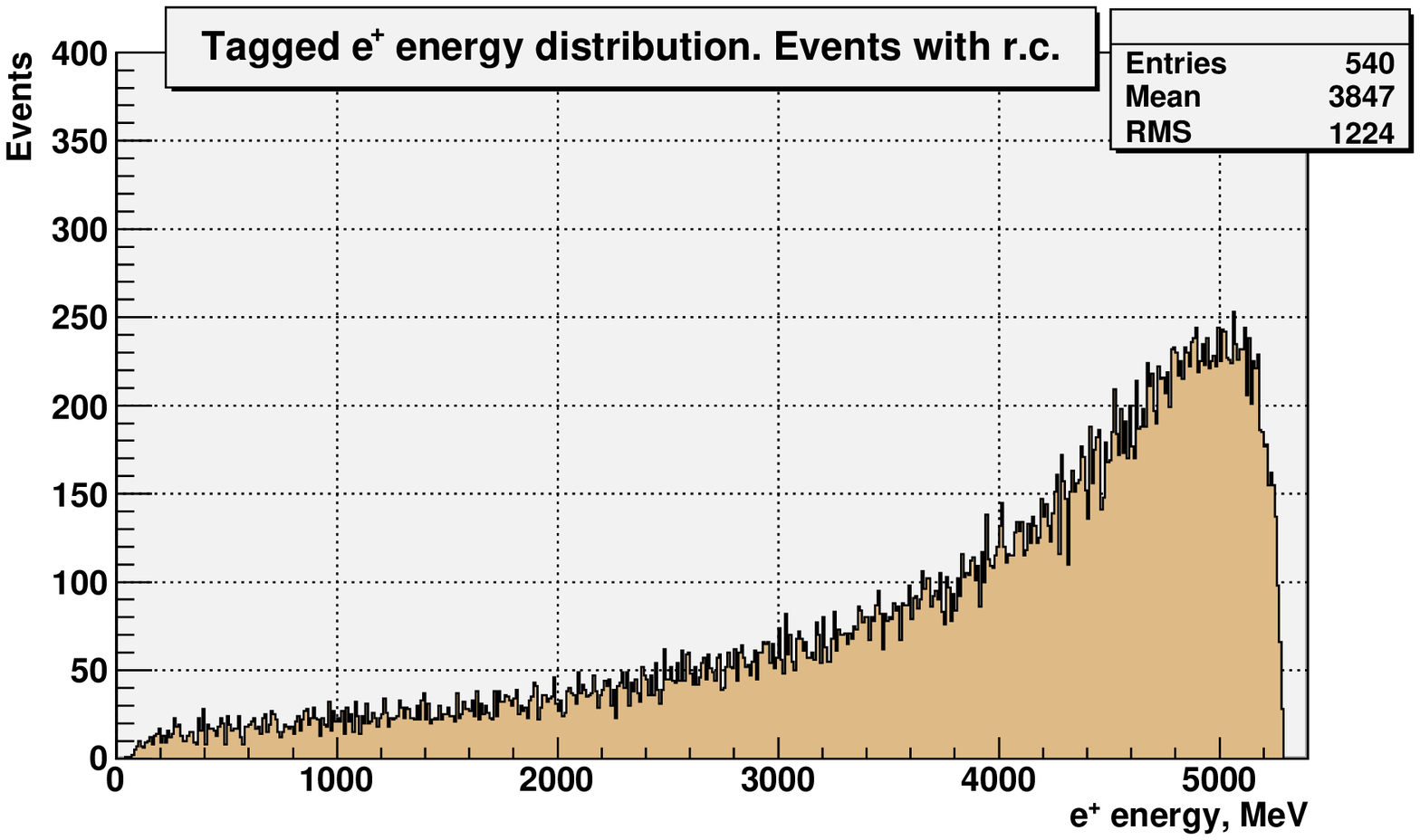}
\caption{\label{fig8}
The energy spectra of tagged electrons from the process
$e^+ e^- \to e^+ e^- \pi^0$ calculated without (left panel) and with
(right panel) radiative correction simulation.}
\end{figure}

The energy spectra of tagged electrons obtained with and without 
radiative-correction simulation are shown in Fig.~\ref{fig8}.
It is seen that emission 
of extra photons significantly changes the shape of this spectrum.

\begin{figure}[hbt!]
\includegraphics[width=.45\textwidth]{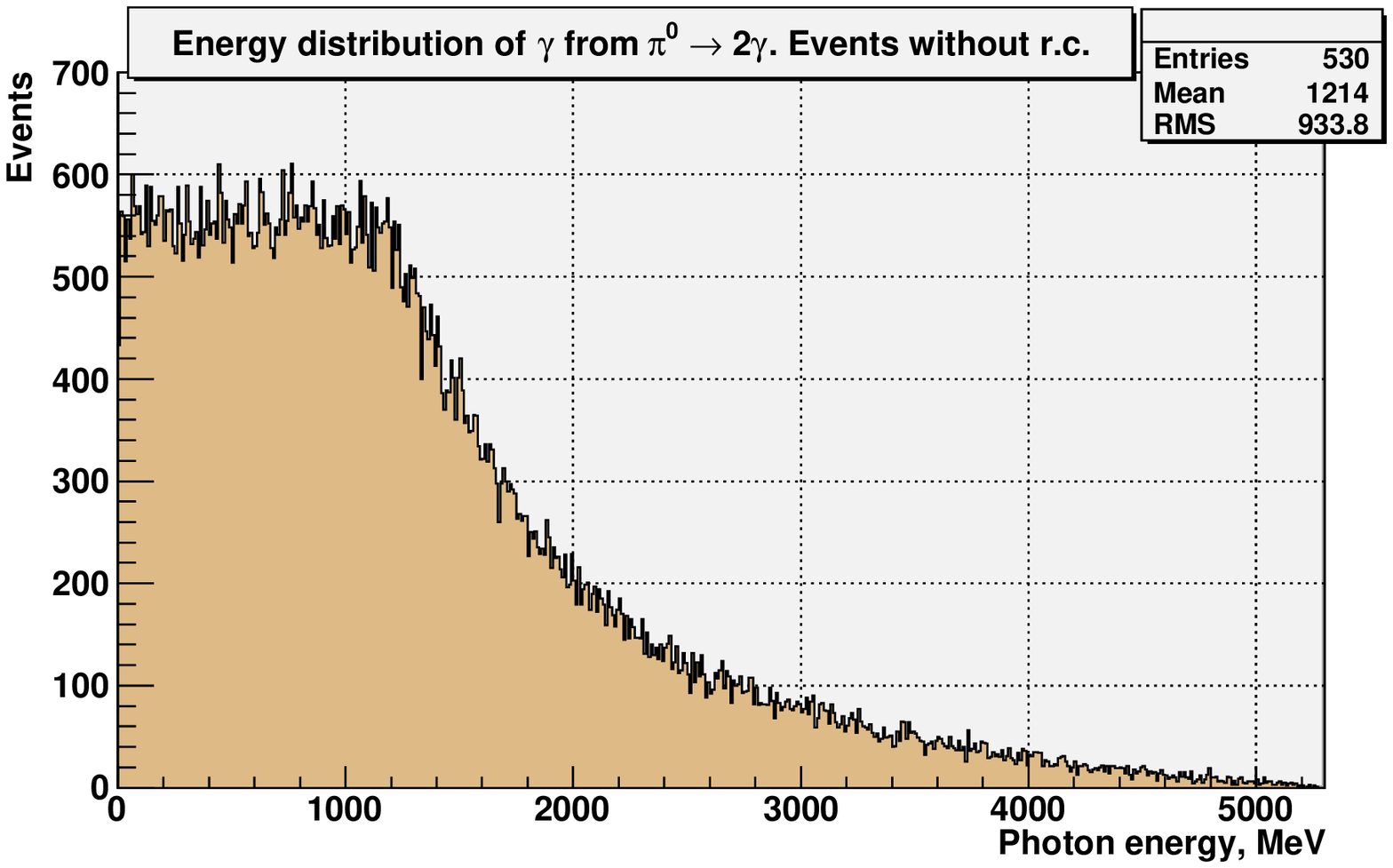}
\includegraphics[width=.45\textwidth]{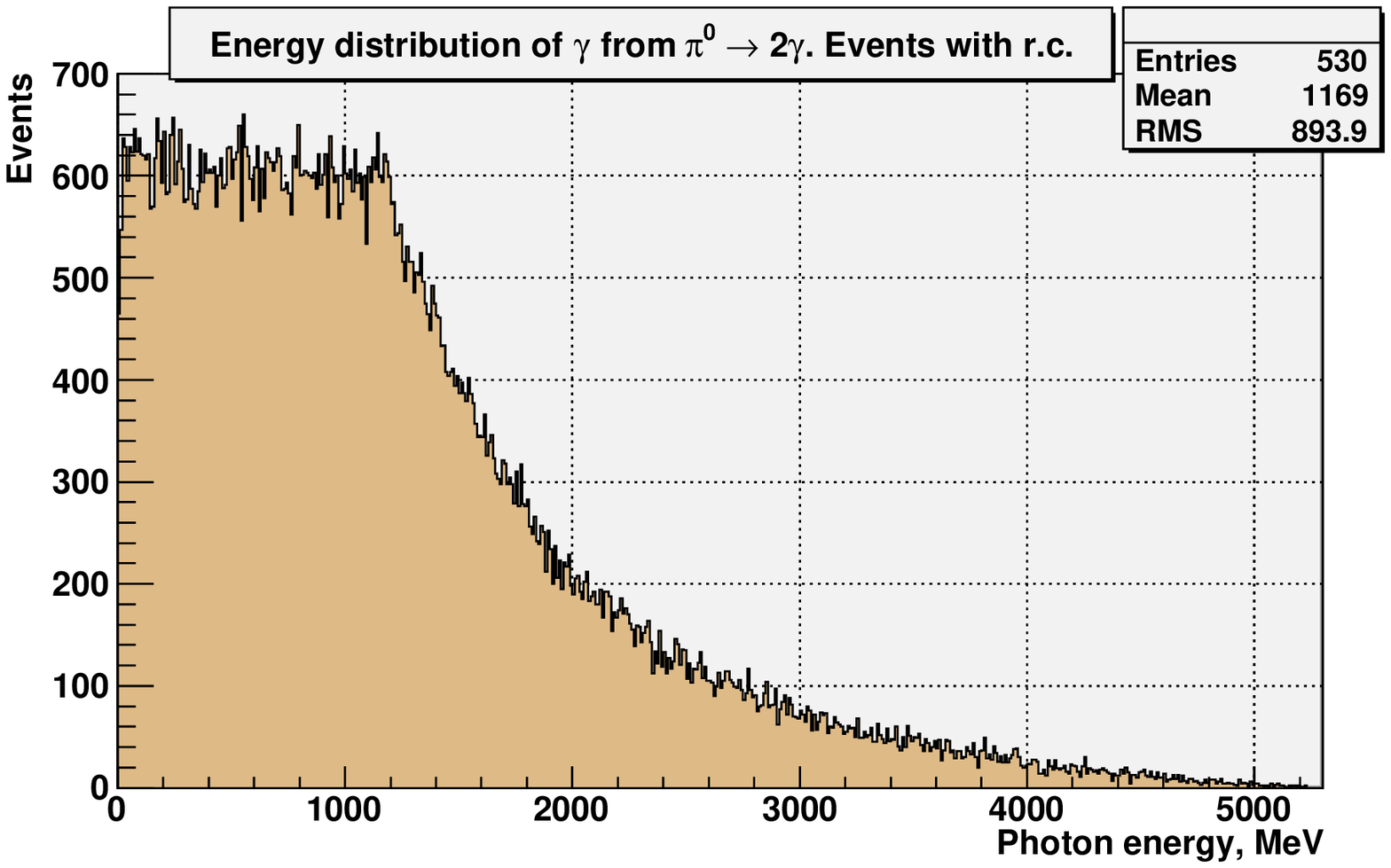}
\caption{\label{fig9}
The energy spectra of photons from the $\pi^0 \to 2\gamma$
decay in the process $e^+ e^- \to e^+ e^- \pi^0$ calculated 
without (left panel)
and with (right panel) radiative correction simulation.}
\end{figure}

\begin{figure}[hbt!]
\includegraphics[width=.45\textwidth]{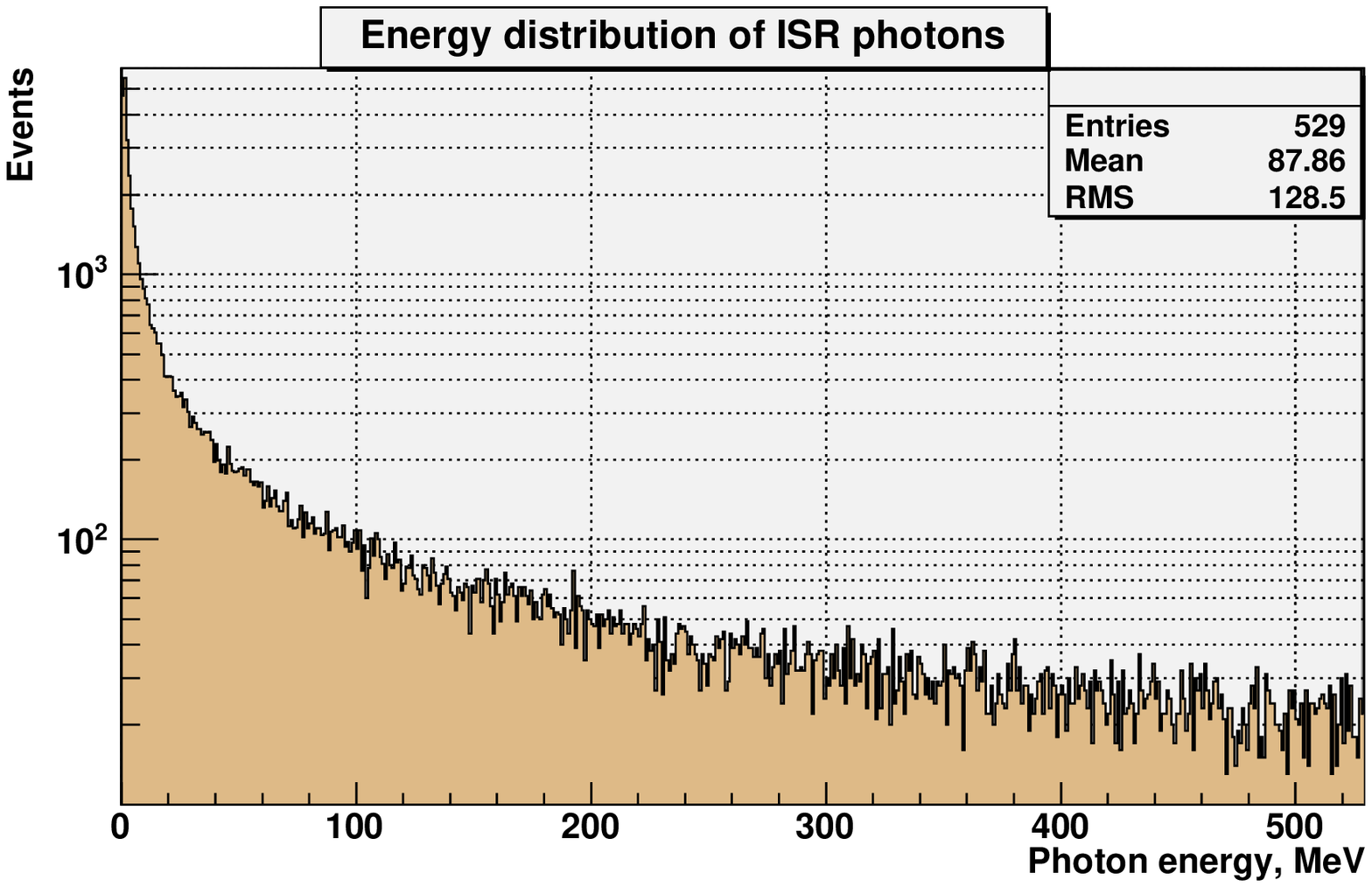}
\includegraphics[width=.45\textwidth]{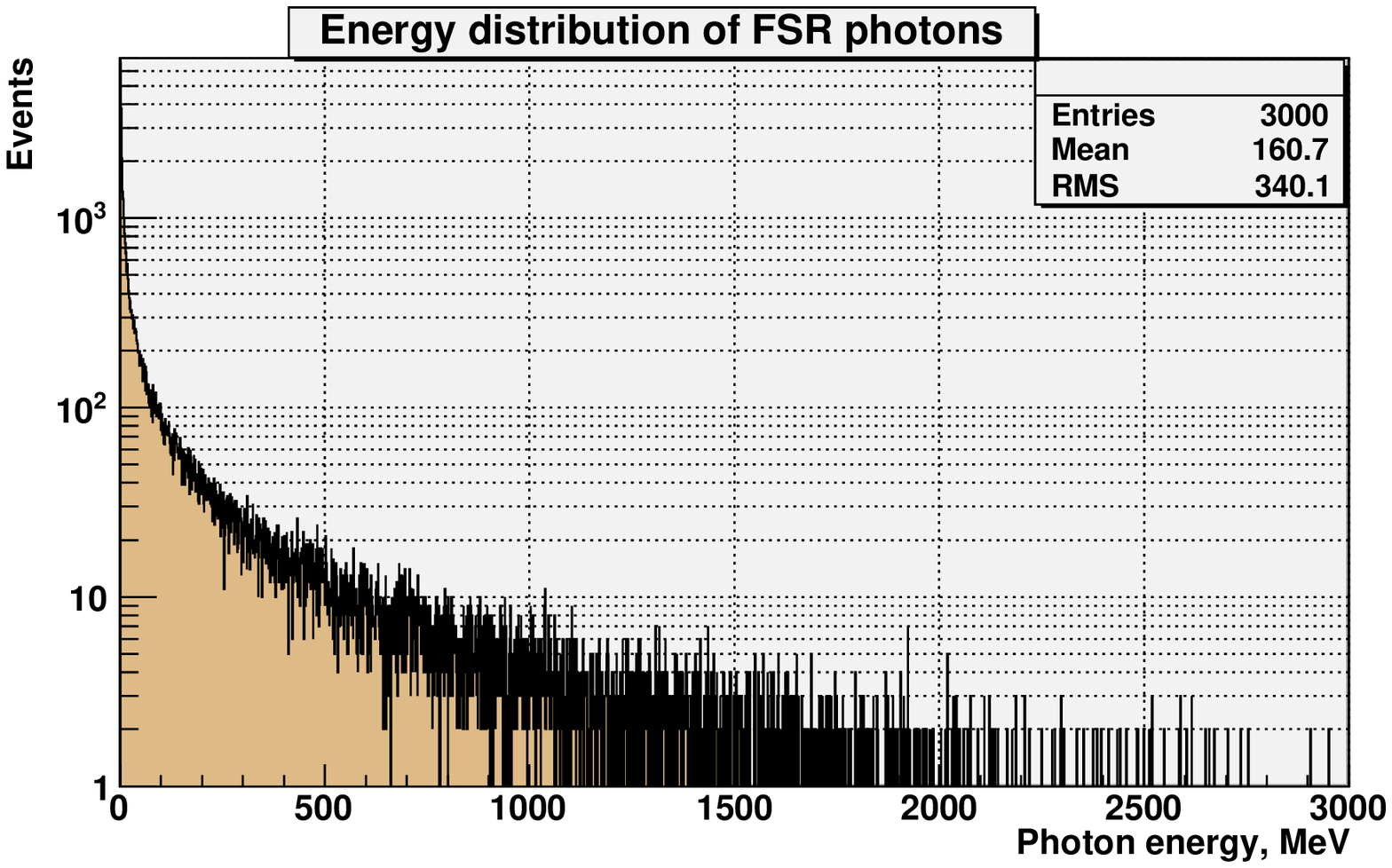}
\caption{\label{fig10}
The energy spectrum of ISR (left  panel) and FSR (right panel) photons
in the simulation of the  process $e^+ e^- \to e^+ e^- \pi^0$.}
\end{figure}

Fig.~\ref{fig9} shows the energy spectra of the
photons from the $\pi^0 \to 2\gamma$ decay in the process
$e^+ e^- \to e^+ e^- \pi^0$. 
The energy spectra of photons
emitted by the initial and final electrons are presented in Fig.~\ref{fig10}. 

\begin{figure}[hbt!]
\includegraphics[width=.45\textwidth]{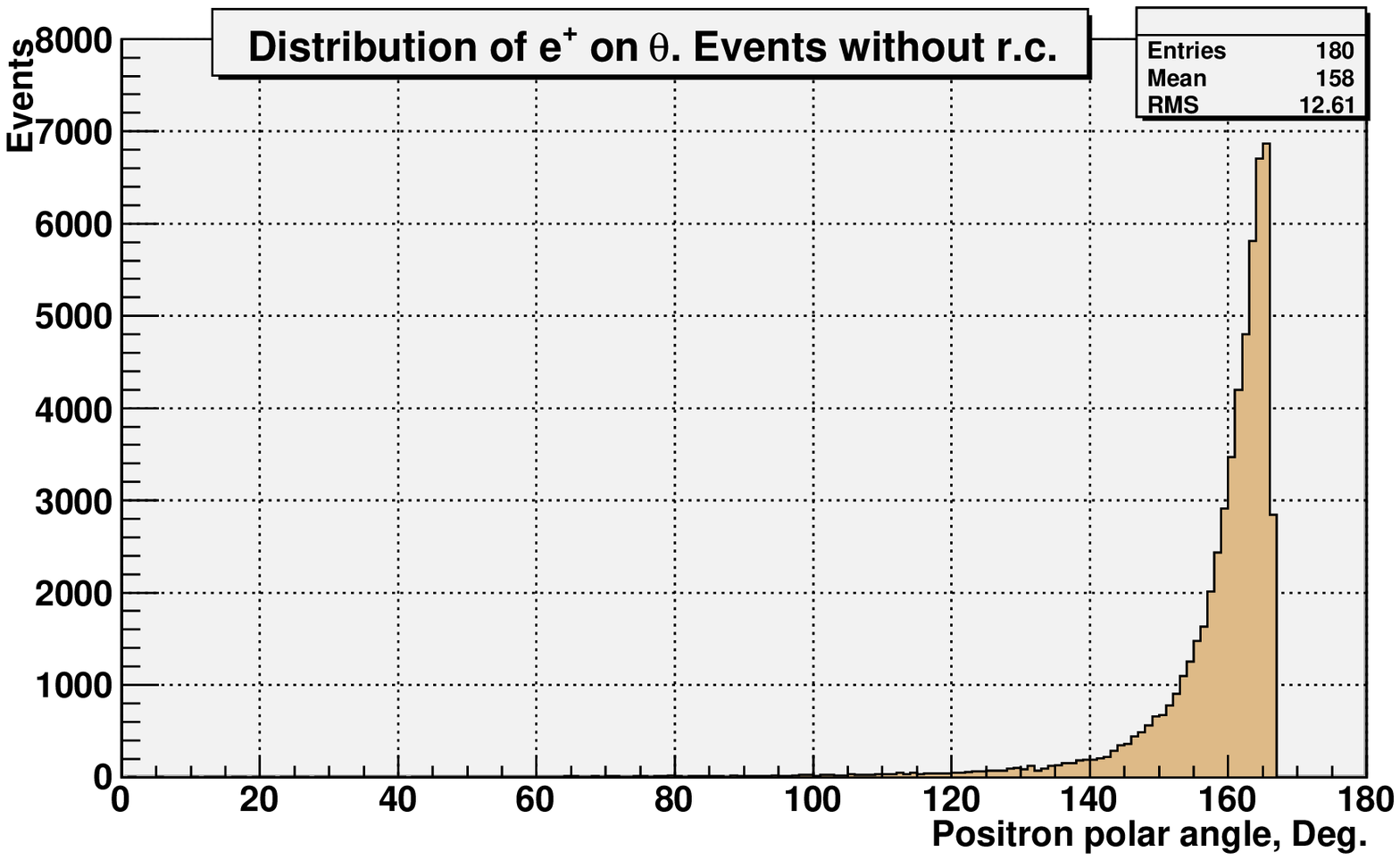}
\includegraphics[width=.45\textwidth]{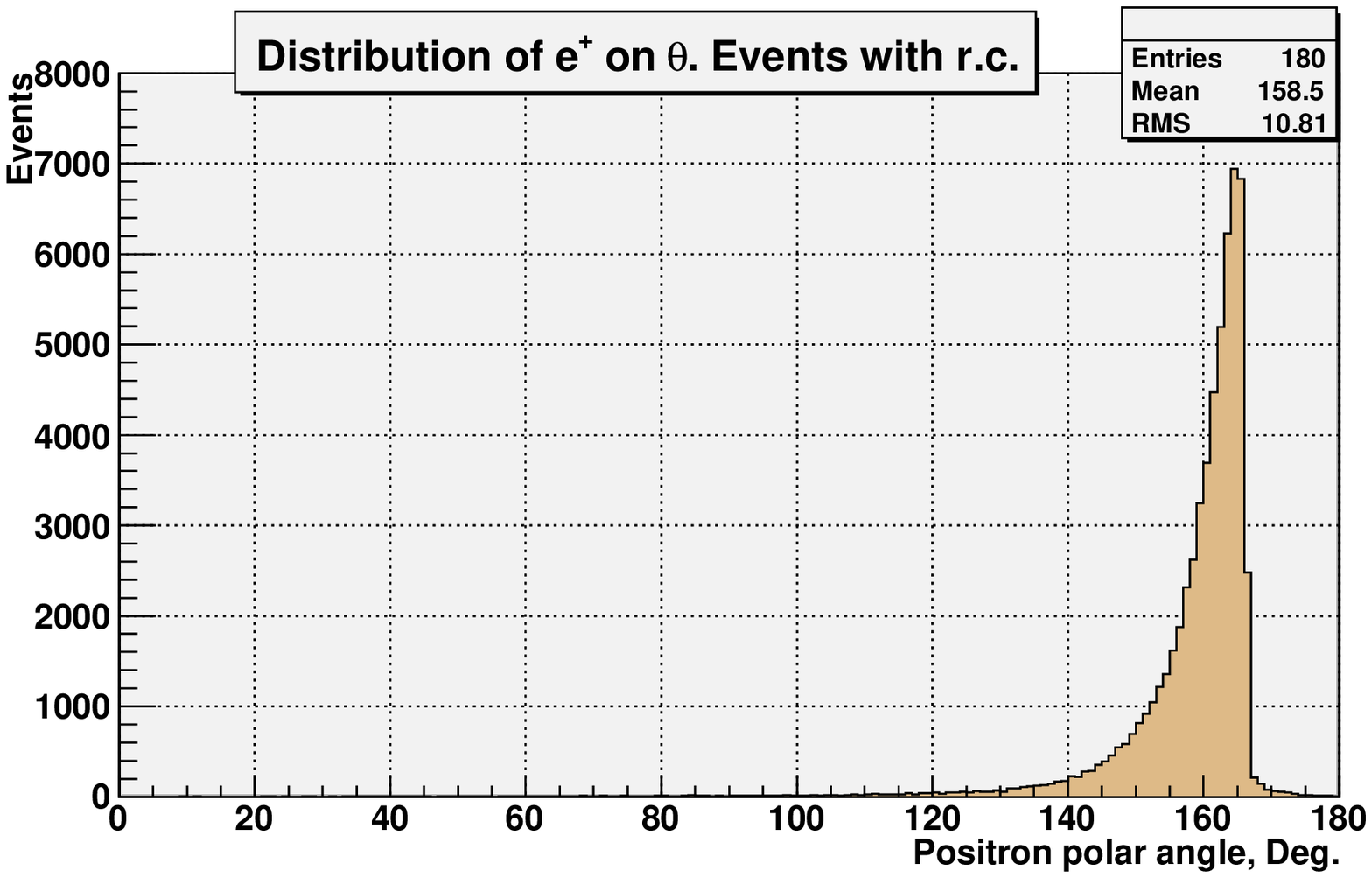}
\caption{\label{fig11}
The polar-angle distributions of tagged electrons 
from the process $e^+ e^- \to e^+ e^- \pi^0$ obtained without (left panel) and with
(right panel) radiative correction simulation.}
\end{figure}

\begin{figure}[hbt!]
\begin{center}
\includegraphics[width=.55\textwidth]{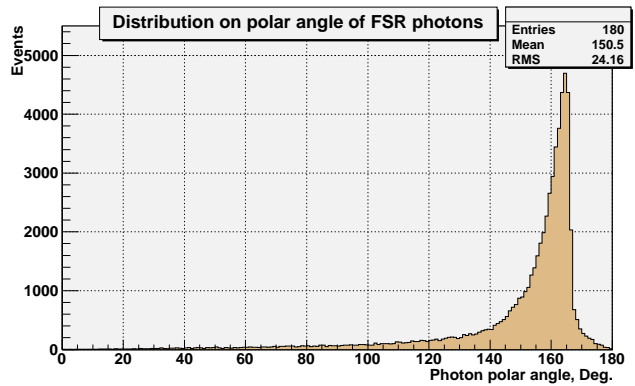}
\caption{\label{fig13}
The polar-angle distributions
of FSR photons from the process $e^+ e^- \to e^+ e^- \pi^0$.}
\end{center}
\end{figure}

The polar-angle distributions of tagged electrons 
is shown in Figs.~\ref{fig11}.
It is seen that at $E_b=5.29$ GeV the
cut $Q^2_{min}>1.5$ GeV$^2$ corresponds to the minimum scattering angle of
about $13^\circ$. This is in agreement with an estimate for small 
scattering angles \linebreak $\theta \approx Q/(E_b \cdot E')^1/2$, where energy 
of the scattered electron $E' \approx E_b$.
\newpage
The polar angle distribution of FSR  photons is shown in Fig.~\ref{fig13}.
Since the FSR photon is emitted predominantly along the tagged-electron
direction, the photon angular distribution is 
very close to that for the electron. 

\begin{figure}[hbt!]
\includegraphics[width=.45\textwidth]{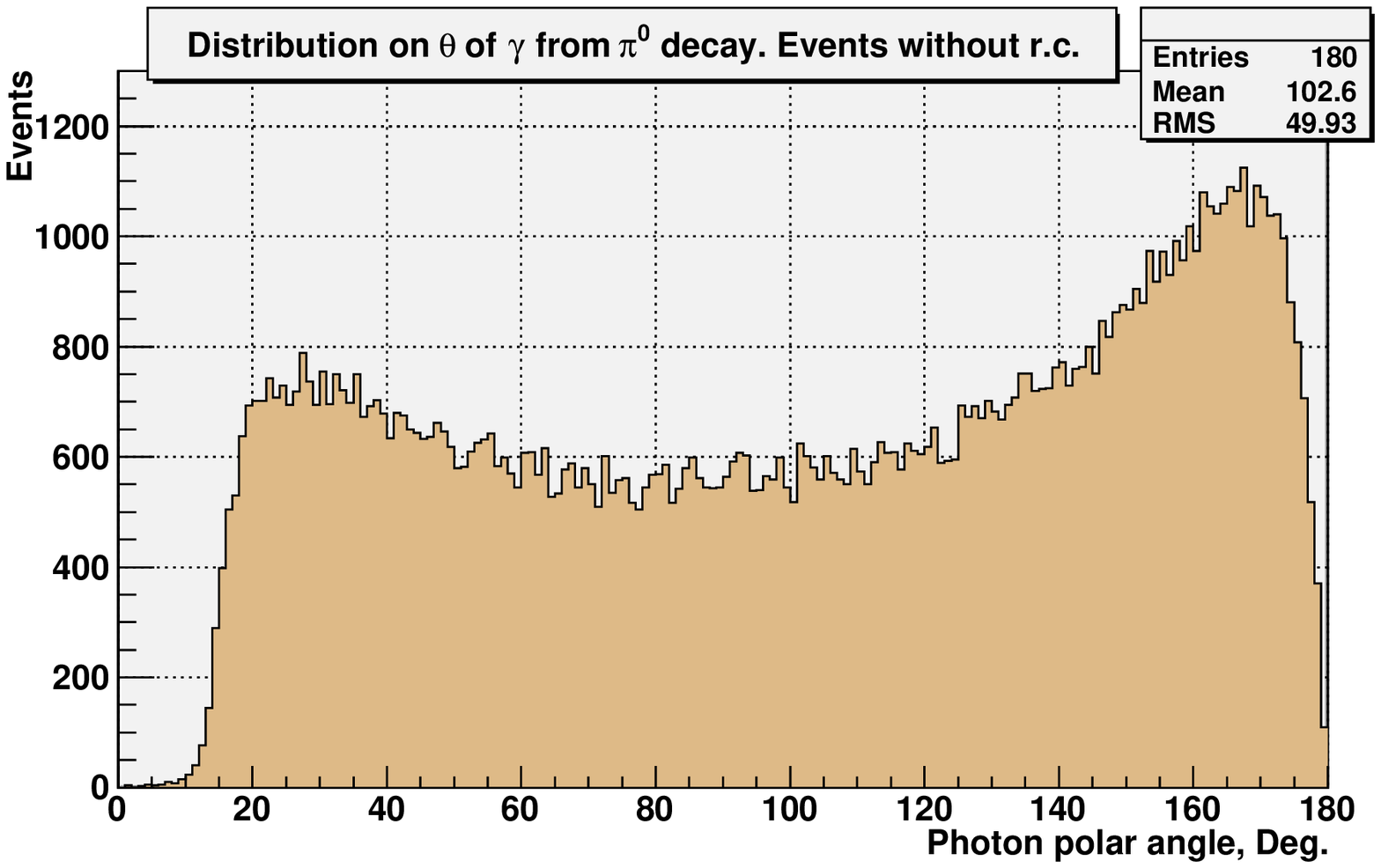}
\includegraphics[width=.45\textwidth]{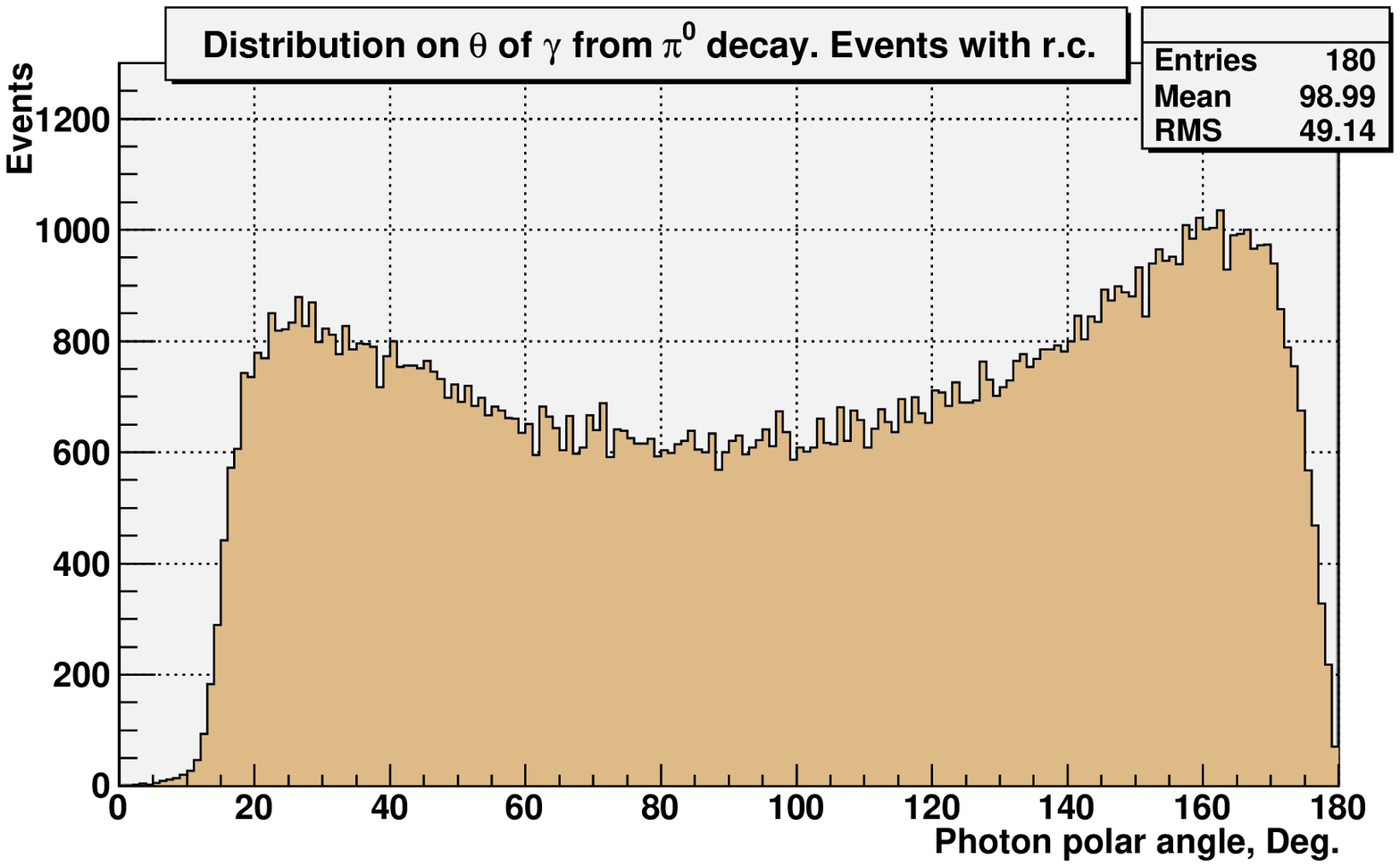}
\caption{\label{fig12}
The polar-angle distributions
of photons from the $\pi^0 \to 2\gamma$
decay in the process $e^+ e^- \to e^+ e^- \pi^0$ obtained without (left
panel)
and with (right panel) radiative correction simulation.}
\end{figure}

The polar-angle distribution of photons
from the $\pi^0 \to 2\gamma$ decay
is shown in Fig.~\ref{fig12}. Photons have wide distribution, which becomes
more uniform with account of radiative corrections.

\begin{figure}[hbt!]
\begin{center}
\includegraphics[width=.7\textwidth]{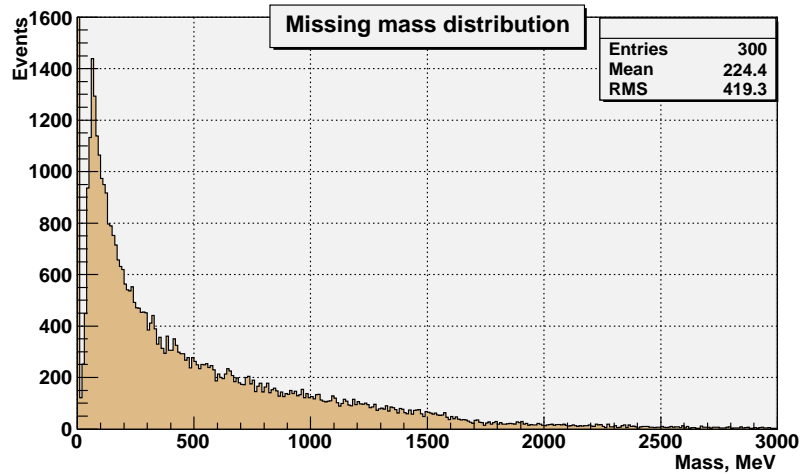}
\caption{\label{fig14}
The missing mass distribution for the simulated $e^+ e^- \to e^+ e^- \pi^0$
single-tag events.}
\end{center}
\end{figure}

In Fig.~\ref{fig14} the distribution of the missing mass 
in the process $e^+ e^- \to e^+ e^- + \pi^0$ is shown.
The missing mass is calculated as 
$\sqrt{(p_1+p_2-k-p_2^\prime)^2}$, i.e. we 
assume that only the tagged electron and the two photon
from $\pi^0$ decay are detected. 
The narrow peak at zero mass contains events (57\% of the total 
number of events),
which  do not have extra ISR or FSR photons. It is seen that emission of the extra
photons leads to significant widening of the missing mass distribution. 

\section {Program components}
\subsection {Common blocks}
\begin{verbatim}
COMMON /GGRSTA/Sum,Es,Sum1,Sum2,Fm,Fm1,NOBR,Nact,Ngt
\end{verbatim}
\vspace*{-5mm}
{\em Purpose:} to collect simulation statistics.\\
\verb+Sum  Real*8    + used for calculation of the total cross section \\
\verb+Es   Real*8    + used for calculation of the cross-section error \\
\verb+Sum1 Real*8    + used for calculation of the average form factor \\
\verb+Sum2 Real*8    + used for calculation of the average radiative correction factor \\
\verb+Fm   Real*8    + maximum weight of event \\
\verb+Fm1  Real*8    + maximum weight of event in FSR simulation \\
\verb+NOBR Integer*4 + number of calls of the generator \\
\verb+Nact Integer*4 + number of the generated events \\
\verb+Ngt  Integer*4 + number of events with the weight greater than Fmax 
(see \verb+/GGRPAR/+) 
\begin{verbatim}
COMMON /GGRPAR/Eb,Rmas,Rwid,Rg,Rm,Fmax,Rmax,Rmin,t1imin,t1imax,
               t2imin,t2imax,Kmin,Fmax1,IR,IMode,KVMDM,ITag,IRad
\end{verbatim}
\vspace*{-5mm}
{ \em Purpose:} simulation parameters.\\
\verb+Eb      Real*8    + beam energy (GeV) \\
\verb+Rmas    Real*8    + meson mass (GeV) \\
\verb+Rwid    Real*8    + meson total width (GeV) \\
\verb+Rg      Real*8    + meson two-photon width (keV) \\
\verb+Rm      Real*8    + meson mass in the current event (GeV) \\
\verb+Fmax    Real*8    + expected maximum weight of event  \\
\verb+Rmax    Real*8    + maximal energy of ISR photon in \verb+Eb+ units \\
\verb+Rmin    Real*8    + minimal energy of ISR photon in \verb+Eb+ units \\
\verb+t1imin  Real*8    + minimal value of $t_1$ (GeV$^2 $) \\
\verb+t1imax  Real*8    + maximal value of $t_1$ (GeV$^2 $) \\
\verb+t2imin  Real*8    + minimal value of $t_2$ (GeV$^2 $) \\
\verb+t2imax  Real*8    + maximal value of $t_2$ (GeV$^2 $) \\
\verb+Kmin    Real*8    + minimal energy of the FSR photon (GeV) \\
\verb+Fmax1   Real*8    + maximum expected weight for FSR simulation \\
\verb+IR      Integer*4 + meson type \\
\verb+Imode   Integer*4 + meson decay mode \\
\verb+KVMDM   Integer*4 + form factor model \\
\verb+ITag    Integer*4 + tagged particle \\
\verb+IRad    Integer*4 + switch for radiative correction calculation  \\
\begin{verbatim}
COMMON /GGRCON/Alpha,PI,EM,mPi0,mPi,mEta,mEtap,mKs,mKc,mRho,mJpsi,
               mUps,BrPi0(2),BrEta(4),BrEtaPrim(4),BrRho,BrTot 
\end{verbatim}
\vspace*{-5mm}
{ \em Purpose:} constants.\\
\verb+Alpha        Real*8 + fine structure constant (1/137.03604) \\
\verb+Pi           Real*8 + $\pi$ (3.14159265) \\
\verb+Em           Real*8 + electron mass (0.00051099891 GeV) \\  
\verb+mPi0         Real*8 + $\pi^0$ mass (0.1349766 GeV) \\
\verb+mPi          Real*8 + $\pi^\pm$ mass (0.13957018 GeV) \\
\verb+mEta         Real*8 + $\eta$ mass (0.547853 GeV) \\
\verb+mEtap        Real*8 + $\eta'$ mass (0.95766 GeV) \\
\verb+mKs          Real*8 + $K_S$ mass (0.497614 GeV) \\
\verb+mKc          Real*8 + $K^\pm$ mass (0.493677 GeV) \\ 
\verb+mRho         Real*8 + $\rho^0$ mass (0.77549 GeV) \\
\verb+mJpsi        Real*8 + $J/\psi$ mass (3.096916 GeV) \\
\verb+mUps         Real*8 + $\Upsilon$ mass (9.4603 GeV) \\
\verb+BrPi0(2)     Real*8 + $\pi^0$ decay branching fractions\\
\verb+BrEta(4)     Real*8 + $\eta$ decay branching fractions\\
\verb+BrEtaPrim(4) Real*8 + $\eta'$ decay branching fractions\\
\verb+BrRho        Real*8 + branching fraction of the decay $\rho^0 \to \pi^+\pi^-$\\
\verb+BrTot        Real*8 + total probability of the decay chain 
\begin{verbatim}
COMMON /GGREV/pPart(4,25),mPart(25),Type(25),Mother(25),Npart 
\end{verbatim}
\vspace*{-5mm}
{ \em Purpose:} final particle parameters  (up to 25 particles).\\
\verb+pPart(1-3,i) Real*8    + momentum of i-th particle (GeV) \\
\verb+pPart(4,i)   Real*8    + energy of i-th particle (GeV) \\
\verb+mPart(i)     Integer*4 + mass of i-th  particle (GeV) \\
\verb+Type(i)      Integer*4 + type of i-th particle \\
\verb+Mother(i)    Integer*4 + index of parent of i-th particle in \verb+/GGREV/+ \\
\verb+Npart        Integer*4 + total number of particles in \verb+/GGREV/+ 

\par\noindent 
In the common block /GGREV/: 1-st and 2-nd particles are the 
scattered electrons, 3-rd particle is the produced resonance,
4-th e.t.c. particles are the ISR photon (if exists),
the FSR photon (if exists), resonance decay products.   

\begin{verbatim}
COMMON /GGRPOL/SETS(7330),SETPOL(7330)
\end{verbatim}
\vspace*{-5mm}
{ \em Purpose:} vacuum polarization correction.\\
\verb+SETS   Real*8 + momentum transfer squared (GeV$^2$) \\
\verb+SETPOL Real*8 + value of the vacuum polarization correction \\
Common blocks for internal use: \verb+/GGRARIP/+, \verb+/GGRFUC/+.

\subsection {Subroutines of the generator}
\noindent
\verb+GGRESRC  + the main subroutine \\
\verb+GGRDEC2G + simulation of resonance decay to 2$\gamma$ \\
\verb+GGRESEND + print out of simulation results \\
\verb+GGRESINI + initialization \\
\verb+GGRETCD  +  simulation of $\eta_c$ decays \\
\verb+GGRETD   + simulation of $\eta$ decays \\
\verb+GGRET1D  + simulation of $\eta'$ decays \\
\verb+GGRET1D1 + simulation of the decays $\eta' \to \pi^+ \pi^- \eta$ and $\pi^0 \pi^0 \eta$ \\
\verb+GGRET1D2 + simulation of the decay $\eta' \to \rho^0 \gamma$ \\
\verb+GGRFSR   + FSR simulation \\
\verb+GGRFVP   + filling the common block /GGRPOL/ \\
\verb+GGRINV   + simulation of the invariants $t_2$, $t_1$, $s_1$, $s_2$ \\
\verb+GGRLMOM  + calculation of the laboratory momenta of the final electrons and meson  \\
\verb+GGRLOR   + Lorentz transformation \\
\verb+GGRPI0D  + simulation of  $\pi^0$ decays \\
\verb+GGRPI0D1 + simulation of the $\pi^0 \to e^+ e^- \gamma $ decay \\
\verb+GGRPREV  + print out of one event \\
\verb+GGRRNDM  + wrapper of a pseudo-random numbers generator \\
\verb+GGRSPC3  + simulation of the three particle phase space 

\subsection {Double-precision functions}
\noindent
\verb+GGRPOLAR + calculation of the vacuum polarization correction \\
\verb+GGRFU    + function used by the subroutine GGRFSR \\
\verb+GGRFVMDM + calculation of the form factor in the vector dominance model

\subsection {Library subroutines}
In the generator we use following functions from
the CERN program library:\\ 
\verb+RANLUX + generation of pseudo-random numbers
uniformly distributed in the interval (0,1);\\
\verb+DZEROX + computing a zero 
of a real-valued function $f(x) $ in the given interval [a, b].    
\section {Summary}
The event generator GGRESRC for simulation of the two-photon process
$e^+ e^- \to e^+ e^-R $, where $R$ is a pseudoscalar meson,
has been developed. The generator allows to efficiently generate 
two-photon events in the single-tag mode, when one of the final
electrons is scattered at a large angle and may be detected. In this
mode simulation of radiative corrections has been implemented
in the generator including extra photon emission from the initial and
final states.

The generator is used for simulation of experiments with the BABAR detector 
on measurements of the photon-meson transition form factors
(see, for example, Refs.~\cite {Bab_pi0,Bab_etac}), and for simulation of
two-photon experiments with the KEDR detector at VEPP-4M collider.

The work is partially supported by the RF Presidential Grant
for \linebreak Sc.~Sch.~NSh-6943.2010.2.

\end {document}